\documentclass[%
 reprint,
 superscriptaddress,
preprint,
 showpacs,preprintnumbers,onecolumn,
 amsmath,amssymb,
 aps,
 floatfix,
]{revtex4-1}

\usepackage[T1]{fontenc} 
\usepackage{mathptmx}
\usepackage{color}

\usepackage{graphicx}
\usepackage{dcolumn}
\usepackage{bm}
\usepackage[colorlinks,breaklinks,
 citecolor=red,linkcolor=red,urlcolor=blue
]{hyperref}

\begin{document}
\bibliographystyle{apsrev4-1}

\title{Topological characterizations of an extended Su-Schrieffer-Heeger model}%

\author{Dizhou Xie}%
\author{Wei Gou}
\author{Teng Xiao}
\affiliation{%
Zhejiang Province Key Laboratory of Quantum Technology and Device, Department of Physics and State Key Laboratory of Modern Optical Instrumentation, Zhejiang University, Hangzhou, Zhejiang, 310027, China
}

\author{Bryce Gadway}
\affiliation{Department of Physics, University of Illinois at Urbana-Champaign, Urbana, IL 61801-3080, USA}
\author{Bo Yan}%
 \email{yanbohang@zju.edu.cn}
\affiliation{%
Zhejiang Province Key Laboratory of Quantum Technology and Device, Department of Physics and State Key Laboratory of Modern Optical Instrumentation, Zhejiang University, Hangzhou, Zhejiang, 310027, China
}
\affiliation{
Collaborative Innovation Centre of Advanced Microstructures, Nanjing University, Nanjing 210093, China
}%

\date{\today}

\begin{abstract}
The Su-Schrieffer-Heeger (SSH) model perhaps is the easiest and the most basic model for topological excitations. Many variations and extensions of the SSH model have been proposed and explored to better understand both fundamental and novel aspects of topological physics. The SSH4 model has been proposed theoretically as an extended SSH model with higher dimension (the internal dimension changes from two to four). It has been proposed that the winding number in this system can be determined through a higher-dimensional extension of the mean chiral displacement measurement, however this has not yet been verified in experiment. Here we report the realization of this model with ultracold atoms in a momentum lattice. We verify the winding number through measurement of the mean chiral displacement in a system with higher internal dimension, we map out the topological phase transition in this system, and we confirm the topological edge state by observation of the quench dynamics when atoms are initially prepared at the system boundary.
\end{abstract}

\maketitle

\section{\label{section1}Introduction}
Topological phases represent an exotic form of matter with geometrical origins. These phases can emerge without any symmetry breaking, which conflicts with the traditional Ginzburg-Landau paradigm. The topological phase is characterized by global properties rather than a local order, making it very robust to certain local perturbations; this topological robustness makes emergent topological excitations a promising candidate for quantum computing. Since the discovery of topological phases in the 1980s, comprehensive studies both in theory and experiment have been carried to create, classify, and comprehend these exotic phases. Various systems across a range of platforms have been engineered to show topological properties, such as the solid \cite{Qi2011,Wang2017}, photonic \cite{Wang2018,Chen2018}, atomic \cite{Stuhl2015,Goldman2016,Meier2016a,meier2018,Cai2018}, acoustic\cite{Xiao2015, Peng2016} and electronic \cite{Ningyuan2015,Zhu2018} systems.

The ultracold atom system provides a powerful tool to study the exotic topological phases, because all the degrees of freedom are precisely controlled. Using optical lattices formed by laser interference, some topological models that are hard to study in other quantum systems can be realized in ultracold atom experiments, such as the famous Haldane model \cite{Jotzu2014}.  The band structure of the lattice can be engineered by choosing different lattice geometries (such as triangular \cite{Becker2010}, kagome \cite{Jo2012}) or different energy bands \cite{Wirth2011,Xu2016}, and this approach of band structure engineering has been essential for introducing topological properties in a cold atom setting. Spin-orbit coupling (SOC) and artificial gauge fields can also be synthetised using light-atom coupling \cite{Lin2011,Wang2012,Wu2016}, and they have been very important for studying the topological insulators. One important advantage of ultracold atom systems is the natural ability to study dynamical processes \cite{Zhang2018a,Wang2017a} which is nearly impossible for other systems such as electronic materials. By suddenly quenching a system into the topological phase of a Hamiltonian, the resulting dynamical processes provide rich information, and can reveal the underlying topology of the system.

Among the wide variety of topological models, the Su-Schrieffer-Heeger (SSH) model is the most basic and one of the most important models in describing band topology in condensed matter physics \cite{Atala2013,Wang2013,Lohse2016,Nakajima2016,Leder2016}. In order to study additional topological physics, various extended models have been proposed. By adding a periodic modulation to the tunneling or the on-site energy, one can study the driven SSH model \cite {Gomez-Leon2013,DalLago2015,Peng2016}. By adding long range tunnelings between different sites, such as the next near neighbor tunneling, one can study the long range hopping SSH model\cite{An2018a,Perez-Gonzalez2018}. By extending the one dimensional model to two chains, one can study the Creutz ladder model \cite{Sun2017,Juenemann2017}.

One direct extension of the SSH model is the so-called SSH4 model \cite{Maffei2018}. By changing the site period of the unit cell from two to four, one can transform the standard SSH model into the considerably richer SSH4 model. As shown in Fig. 1(a), the tunneling rates between pairs of sites repeat every four lattice sites, and we label the fundamental tunneling terms as $\{t_a,t_b,t_c,t_d\}$. We additionally label the sublattice site positions within each unit cell as $\{A_1,B_1,A_2,B_2\}$. In the situation where $t_a = t_c$ and $t_b = t_d$, the SSH4 model reduces to the common SSH model. 
For a SSH4 model with infinite sites, the topological phase is determined by the tunneling ratio $\gamma=bd/ac$. If $\gamma>1$, it is topologically nontrivial and can hold topological, zero-energy edge states. If $\gamma<1$, it is topological trivial and no edge state exists. In Fig. 1(b), we plot the eigenenergy of the SSH4 system with six unit cells versus $1/\gamma$ with six unit cells. Here we choose $t_b=t_c=t_d=2\pi\times$ 1kHz and vary $t_a$.  There exist states at zero energy when $1/\gamma<1$, and these states, the topological edge states, can be shown to exist at the boundary of the system. Figure 1(c) shows one typical population distribution for the edge state when $1/\gamma=1/2$. The population is mainly distributed within the first unit cell.

One of the interesting features of SSH4 model lies in the much wider parameter space. The larger parameter-space of the SSH4 model with an enhanced unit cell is useful for studying topological properties with higher dimensions  \cite{lustig2019, Celi2014, Stuhl2015} and potentially also a higher winding number \cite{Xiao2018} . When periodical driving is applied to the SSH4 system, an effectively higher dimensional quantum model can be realized. Reference \cite{Maffei2018} proposed an example of forming a discrete-time quantum walk with effectively four dimensions by driving the SSH4 model. interesting feature of the SSH4 model relates to the methods for detecting the winding number in such systems. The mean chiral displacement has been proposed and experimentally verified as an observable that reveals the winding number in the standard SSH model \cite{Cardano2017, meier2018}. Reference \cite{Maffei2018} provides a generalized description for how this observable can be extended to higher dimensional systems, however it has yet to be verified in experiment. In this manuscript, we have realized the higher-dimensional (in the internal dimension of the unit cell) SSH4 model with ultracold atoms in momentum space, which would be quite difficult to engineer in real space. We have verified the utility of the mean chiral displacement measurement in this higher dimensional model, using it to map out the topological phases of the SSH4 system.

\section{\label{section2}Results}
In experiment, we use ultracold atoms in a momentum lattice to realize the SSH4 model. Such quantum simulator is a versatile platform for studying topological models\cite{Meier2016}. We first produce a $^{87}$Rb Bose-Einstein condensate (BEC) in a crossed dipole trap \cite{Xie2018}. The BEC contains about $6\times 10^4$ atoms, and the trap frequencies are approximately $2\pi\times(90, 90, 90)\mathrm{Hz}$. We use an additional $1064$ nm beam to Raman couple the different momentum states. As shown in Fig. 2(a), the incoming beam ($\omega_+$) passes through the atomic BEC and is then reflected back upon itself, in the antiparallel direction. It goes through two AOMs before again passing through the cloud of atoms. One AOM is driven by $f_0=100\mathrm{MHz}$ radio frequency (RF) and the $+1$ order is chosen, while for the second one $f=f_0+\Sigma f_n$ and the $-1$ order is chosen. Atoms can absorb a photon from one direction ($\omega_+$) and undergo stimulated emission of a photon in to the opposite direction ($\omega_-$). Thus, a total two-photon recoil momentum $p_0=2\hbar k$ is transferred to the atom. As shown in Fig. 2(b), each pair of the Raman beams $\{\omega_+ \oplus \omega_-^n\}$ couple the momentum states from $\{np_0\}$ to $\{(n+1)p_0\}$. Hereafter, we label the momentum state with momentum $np_0$ as $|n\rangle$. The momentum states $\{...,-2,-1,0,1,2,...\}$ form an effective lattice structure in the momentum space. $f_n$ is set to be $(2n+1)\times4E_r$ ($E_r=h\times2.03\mathrm{kHz}$) to resonantly couple different momentum states from $|n\rangle$ to $|n+1\rangle$. In our experiment, we detect the atomic distribution by suddenly turning off all of the laser beams and allowing atoms to fall in free space for $20$~ms, such that atoms in different momentum states evolve to different positions. Then we take an absorption image and count the number of atoms in the different momentum states.

This Raman coupled momentum lattice structure can be described by a simple tight-binding model. With the rotating wave approximation, the Hamiltonian is \cite{Gadway2015}
\begin{equation}
H=\sum\limits_{n} t_n[e^{i\phi_n} (|\widehat{\psi}_{n+1}\rangle\langle\widehat{\psi}_n|+H.c.)],
\end{equation}
where $\phi_n$ are the relative phases for different tunnelings, which are determined by the relative phases between lasers with different frequencies. In our case, all these phases are set to be zero. $t_n$ are the tunneling rates, determined by the Raman coupling strength of each Raman beam pair, $t_n=\Omega_+\Omega_n/{4\Delta}$. They can be adjusted by changing the laser intensity of each discrete frequency component. If the $t_n$ are set to take different values for the alternating tunneling links, the Hamiltonian can be mapped to the standard SSH model. If $t_n$ are set to be periodic over every four sites, this system realizes the SSH4 model.

For one dimensional chiral models, the winding number $\upsilon$ is an important topological invariant used to characterize the topological phase. The number of edge states on each edge is $|\upsilon|$. It can be measured by detecting the phase of a particle across the Brillouin zone \cite{Atala2013}, and can also be measured with a quench dynamics \cite{Maffei2018}. The mean chiral displacement was recently introduced to measure the winding number and has been conducted in the photonic system \cite{Cardano2017} and cold atom system \cite{meier2018} for the SSH model, and it is quite insensitive to disorder. The mean chiral displacement is defined as
\begin{equation}
C(t)=\langle\Gamma m\rangle;
\end{equation}
which quantifies the relative shift between the projections of the state onto the eigenstates of the chiral operator $\Gamma$. Here $\hat{m}$ is the unit cell operator.
The dynamics of $C(t)$ converge to the winding number $\upsilon$ for the long time dynamics for initial states beginning within one unit cell following \cite{Cardano2017,meier2018}
\begin{equation}
\upsilon=\textrm{Tr}(\widehat{\Gamma m});
\end{equation}
For the SSH4 model, we can choose the unit cell basis $\{A_1, B_1, A_2, B_2\}$, so the total mean chiral displacement operator takes the form $\widehat{\Gamma m}=diag(...,1,-1,1,-1,2,-2,2,-2,...)$.  At higher dimension, in order to use the mean chiral displacement to measure the winding number, we need to choose an orthogonal and complete basis of a given sub-lattice \cite{Maffei2018}.  In our case, we prepare the initial state at two orthogonal states (1,0,0,0) and (0,0,1,0), and measure the mean chiral displacement $C_1(t)$ and $C_3(t)$ respectively. Then we sum these two measures $C_{total}(t)=C_1(t)+C_3(t)$ to get the total mean chiral displacement which converges to the winding number $\upsilon$ at long evolution times.

Figure 3 shows the experimental results of the mean chiral displacement measurement. The RF driver of the AOM includes 24 discrete frequency teeth, so the unit cell number of this SSH4 model is $N=6$. The initial state is prepared within the central unit cell. By suddenly turning on the Raman couplings, the dynamics of the evolution is recorded and the $C_{total}(t)=C_1(t)+C_3(t)$ is extracted.  To enter the different topological phases, we choose the tunneling about $t_a=t_c=t_d=2\pi\times 1\mathrm{kHz}$, and vary the ratio $t_b/t_a$.  The left inset shows $C_{total}(t)$ when $\gamma<1$, it oscillates around zero. The right inset shows $C_{total}(t)$ when $\gamma>1$, which rises up and then oscillates around one. The dynamics of $C_{total}(t)$ show completely different behaviors in these two regimes. The blue curves are the ideal numerical simulations according to Eq. (1), which capture the main features of the experimental data. We extract the averaged value and then plot them versus different $\gamma$, as shown in Fig. 3. The error bars of the data points mainly stem from uncertainty due to the presence of a thermal fraction (about 10\% of the total population in our experiments). The red curve is the theory prediction for an infinite number of unit cell $N\rightarrow\infty$. When N is a large number, the topological phase transition is sharp in the thermodynamic sense.  The blue dash curve is a numerical simulation with our experimental parameters for N=6, in which case one finds a smooth transition between topologically trivial and topologically nontrivial phases. Our data agree with the theory quite well.

In addition to measuring the bulk topology through the mean chiral displacement, we also directly detect the boundary signatures of the topology through quench dynamics at the edge of our SSH4 model. Figure 4 shows the typical experimental results of edge quench dynamics. The BEC is prepared at the $|0\rangle$ state, and then we turn on the Raman coupling and let the BEC evolve under the SSH4 Hamiltonian. The parameters are chosen as  $t_a=t_c=t_d=2\pi\times0.5\mathrm{kHz}$ and $t_b=2\pi\times1\mathrm{kHz}$. Because $\gamma>1$, the edge states should exist under these conditions.  For Fig. 4(a) and (b), the Raman beams are designed to couple the momentum states of $n=\{0, 1, 2, ..., 19\}$. So, the initial state is prepared at the edge, which is not exactly the eigenstate of the edge state, but has big overlap with its wavefunction. In this case, we will expect population to continuously remain at the $|0\rangle$ position. Figure 4(a) displays the experimental data. The different positions along the x direction relate to the different discrete momentum states, which are labeled at the top of the picture. The time step of our measurements is $50\mu s$. We see that the population in the $|0\rangle$ state remains dominate over time, as expected. Fig. 4(b) is the theoretical simulation with $t_a=t_c=t_d=2\pi\times0.5\mathrm{kHz}$ and $t_b=2\pi\times1\mathrm{kHz}$. The experimental and theoretically simulated dynamics display good agreement. For initial preparations of a bulk state, the system shows distinct dynamical behavior. For Fig. 4(c) and (d), the Raman coupling is designed to couple $n=\{-11, ..., -1, 0, 1, ..., 12\}$ states, such that the initial BEC state is in the Bulk. The population in $|0\rangle$ can be seen to nearly vanish, even at these relatively short times. Figure 4(c) is the experimental data and Fig. 4(d) is the theoretical simulation with $t_a=t_c=t_d=2\pi\times0.5\mathrm{kHz}$ and $t_b=2\pi\times1\mathrm{kHz}$. Again, we find generally good qualitative agreement between the theory and experiment in this case of bulk injection. For both the edge and bulk injection cases, we observe that the experiment and theory are in good agreement at short time, but begin to deviate at longer times. Some possible sources of this deviation, deserving of future investigation, include Raman laser phase noise, the expansion and separation of the momentum states, and atomic interactions.

\section{\label{section7}Discussion}

To conclude, we have experimentally realized a new kind of extended SSH model, the SSH4 model, in a momentum lattice with ultracold atoms. We have measured the bulk topological properties of this system through quench dynamics. We have measured the winding number by means of the mean chiral displacement, and the phase transition is mapped out. We have found quite excellent agreement between the experimental data and the mean chiral displacement theory, which shows a robustness of this dynamical topological observable even in the presence of effects limiting the long-time dynamics. We attribute this to the insensitivity of the mean chiral displacement to disorders as confirmed in \cite{Cardano2017,meier2018}. Our result is the first experimental demonstration of the mean chiral displacement predicted at higher dimension. In addition to our measurements of the bulk topology, we confirm these results by directly observing evidence for topological boundary states at the edge of our SSH4 lattice.

\section{\label{section2}Methods}

The experimental setup is shown in Fig. 5. Two dipole trap (DT) beams are used as the final trap for our BEC. The waist size of the two DT beams are about 70~$\mu$m. The trap frequencies of our DT are roughly $\omega=2\pi\times(90, 90, 90)\mathrm{Hz}$. There are roughly $6\times 10^4$ atoms in our BEC. The Raman beams make roughly a 7 degree angle with the DT2 beam, and have a size of roughly 230~$\mu$m at the position of the BEC. After the chamber, the Raman beam passes through two AOMs and returns to the BEC position.

In order to reduce the momentum spread along Raman direction, we reduce the power of the DT1 laser and increase the Raman beam incoming powers at the same time. The trap frequency along Raman direction becomes about $2\pi\times 40\mathrm{Hz}$. The ramp time is 0.4s and additional 0.1 s allows for rethermalization of the atoms to avoid breathing or sloshing. The incoming Raman beam is kept on and the intensity is stabilized. The Raman coupling is pulsed on by turning on the two AOMs following the BEC and allowing for the retroreflected beam to be turned on, as shown in the left side of Fig. 5.

After the Raman pulse, all the dipole trap beams and the Raman beams are turned off. Atoms fall freely in space. We take an absorption image after 20ms. At this time, atoms with different momentums will occupy different positions, as shown in Fig. 6(a). In order to count the number of atoms in different momentum states, we sum the image along the top-down direction, and then fit it with a multi Gaussian function. The size of each Gaussian peaks is set to be the same, and the distance between nearest peaks is set to be equal. In such constrained fits, the peak value of each Gaussian function is proportional to the atom number at the different momentum states.

\section{\label{section2}Data availability}
The datasets generated during and/or analysed during the current study are available from the corresponding author on reasonable request.

\section{Author contribution}
B. Y. and B. G. proposed the idea. D. -Z. X, W. G and T. X performed the experiment and completed the data analysis. B. Y. supervised the project. All authors contributed for the writing of the manuscript.

\section{Competing Interests}
The authors declare that there are no competing interests.

\begin{acknowledgments}
We acknowledge the support from the National Key R$\&$D Program of China under Grant No.2018YFA0307200, National Natural Science Foundation of China under Grant No. 91736209, National Natural Science Foundation of China under Grant No. 91636104, Natural Science Foundation of Zhejiang province under Grant No. LZ18A040001, and the Fundamental Research Funds for the Central Universities, B. G. acknowledges support from the National Science Foundation under grant No. 1707731.
\end{acknowledgments}

\bibliographystyle{naturemag}

\begin{figure}[]
\includegraphics[width= 0.47\textwidth]{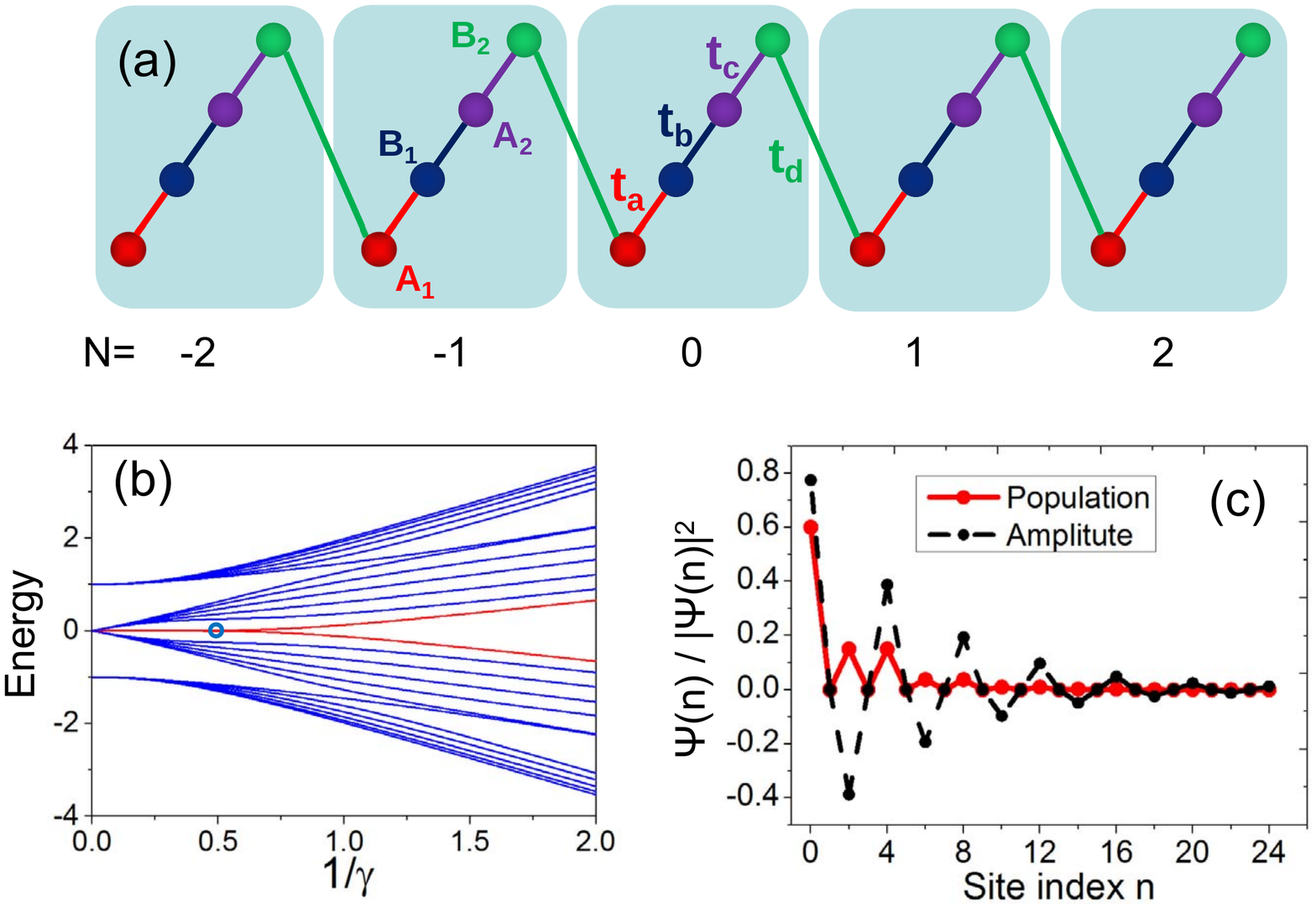}
\caption{\label{figure1}(Color online) (a) The diagrammatic sketch of the SSH4 model. tunneling terms are periodic with a unit cell of 4 sites. (b) The eigenenergy of each eigenstate for different $\gamma$ with the SSH4 model for N=6.  $t_b=t_c=t_d=2\pi\times$1kHz, and $t_a$ varies. The red lines show that there are eigenstates with zero energy when $1/\gamma<1$, which correspond to the existence of topological edge states. And when $1/\gamma>1$, there are no zero-energy eigenstates. For large system (large N number), a sharp phase transition happens at $\gamma=1$. (c) The probability amplitude and the population distribution for an edge state when $1/\gamma=0.5$, as circled in (b).}
\end{figure}

\begin{figure}[]
\includegraphics[width= 0.4\textwidth]{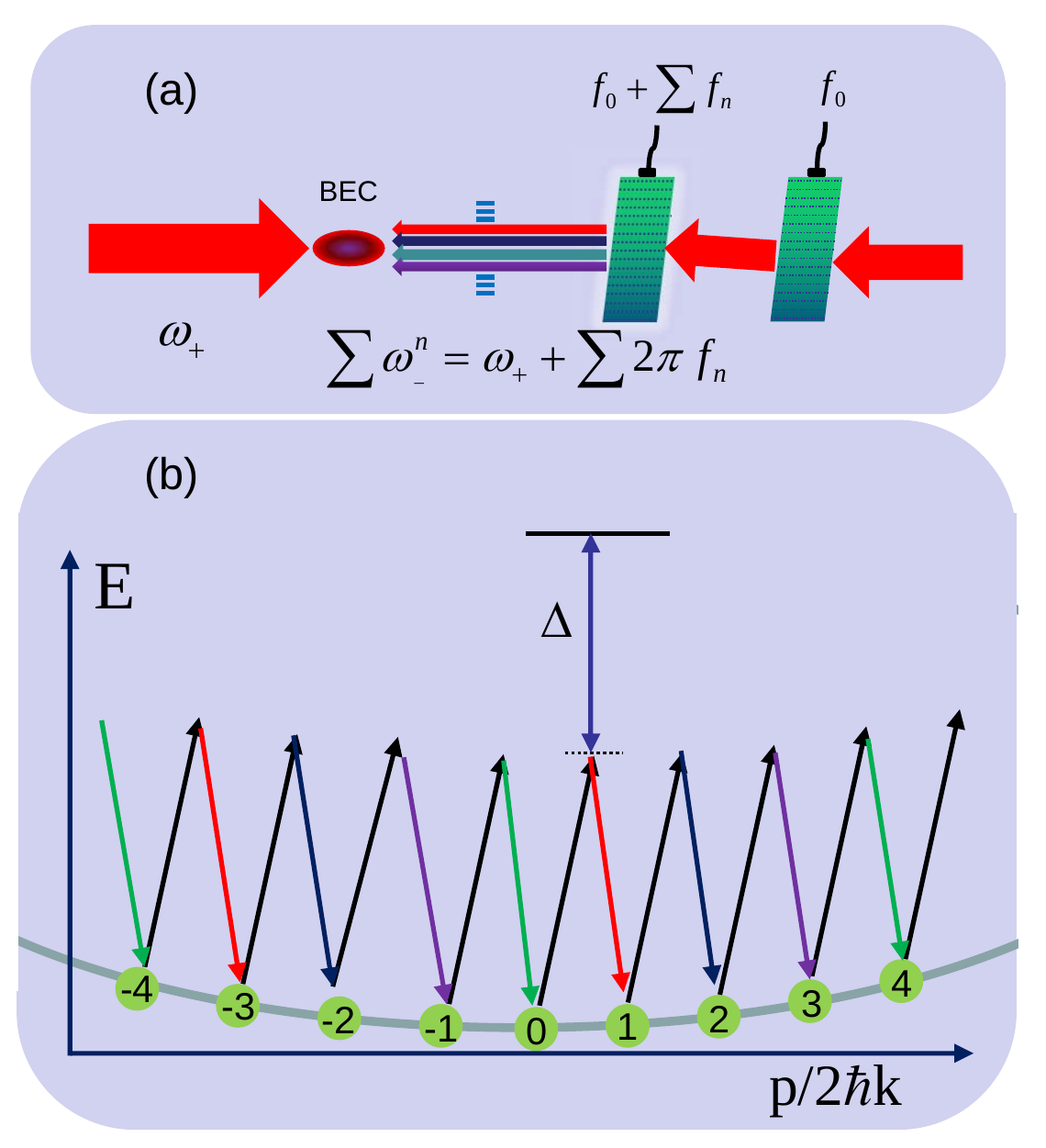}
\caption{\label{figure1}(Color online) (a) The experimental setup for multi-frequency Raman coupling. The incoming Raman beam passes the BEC, then passes two AOMs, and finally propagates back to the BEC. One AOM is modulated with a frequency $f_0$, and the other one is modulated with the multi-frequency signal $f_0+\sum f_n$. One AOM is operated in the positive first order while the other is operated in the negative first order. So the two Raman beams on BEC differ in frequency by $\sum f_n$. (b) An illustration of the Raman processes driven by the pair of Raman beams.}
\end{figure}

\begin{figure}[]
\includegraphics[width= 0.4\textwidth]{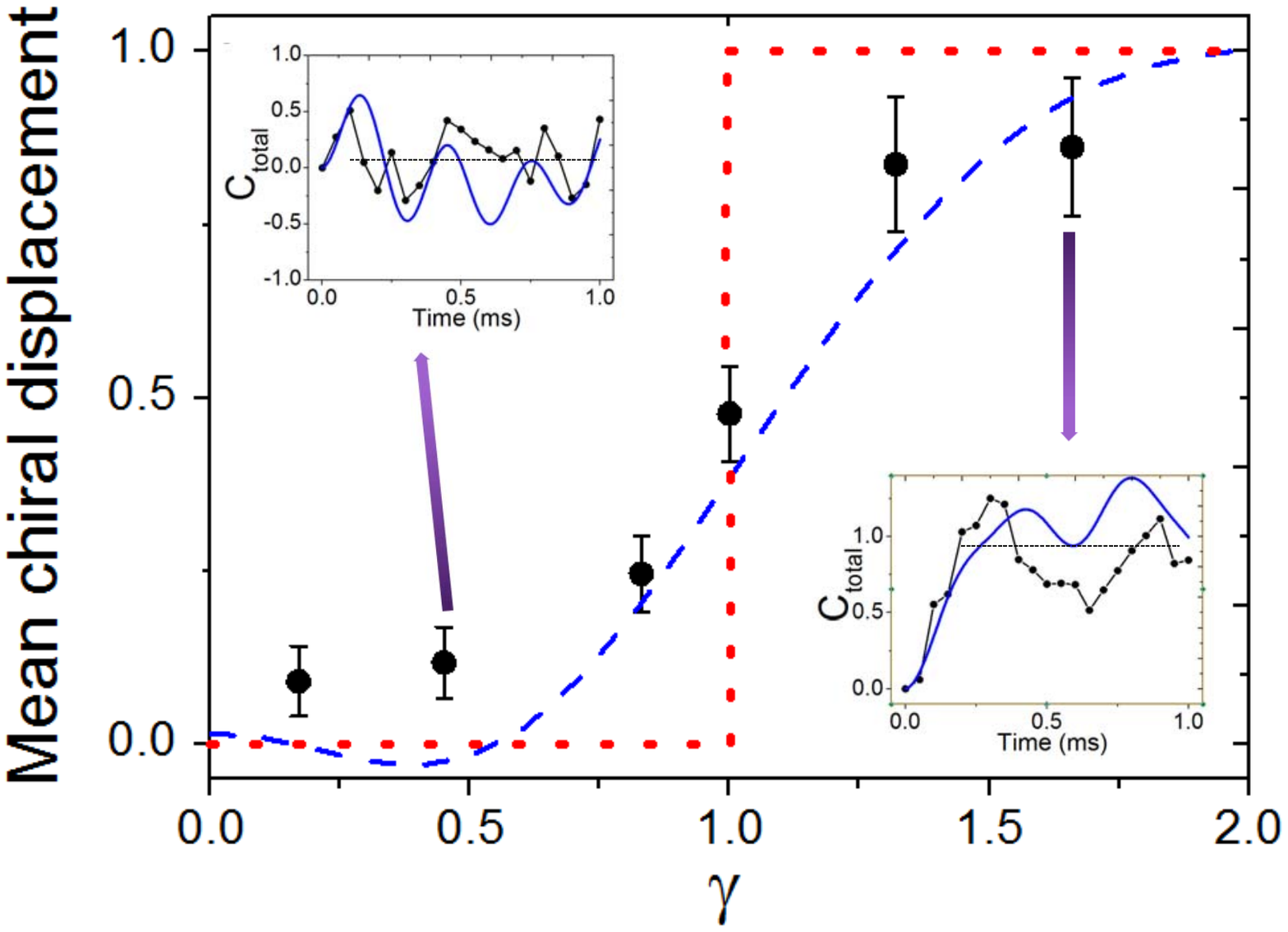}
\caption{\label{figure1}(Color online) The measured total mean chiral displacement versus the tunneling ratio $\gamma$. The black dots are experimental data. The blue dash line is the theoretical simulation with N=6 unit cells. The red dashed line is the theoretical simulation with $N\rightarrow\infty$, where a sharp phase transition occurs at $\gamma=1$. The insets show the two date sets for $\gamma<1$ and $\gamma>1$ respectively. For these inserts, the dots are the experimental data and blue lines are the simulations. Error bars are defined as the standard deviation.}
\end{figure}

\begin{figure}[]
\includegraphics[width= 0.47\textwidth]{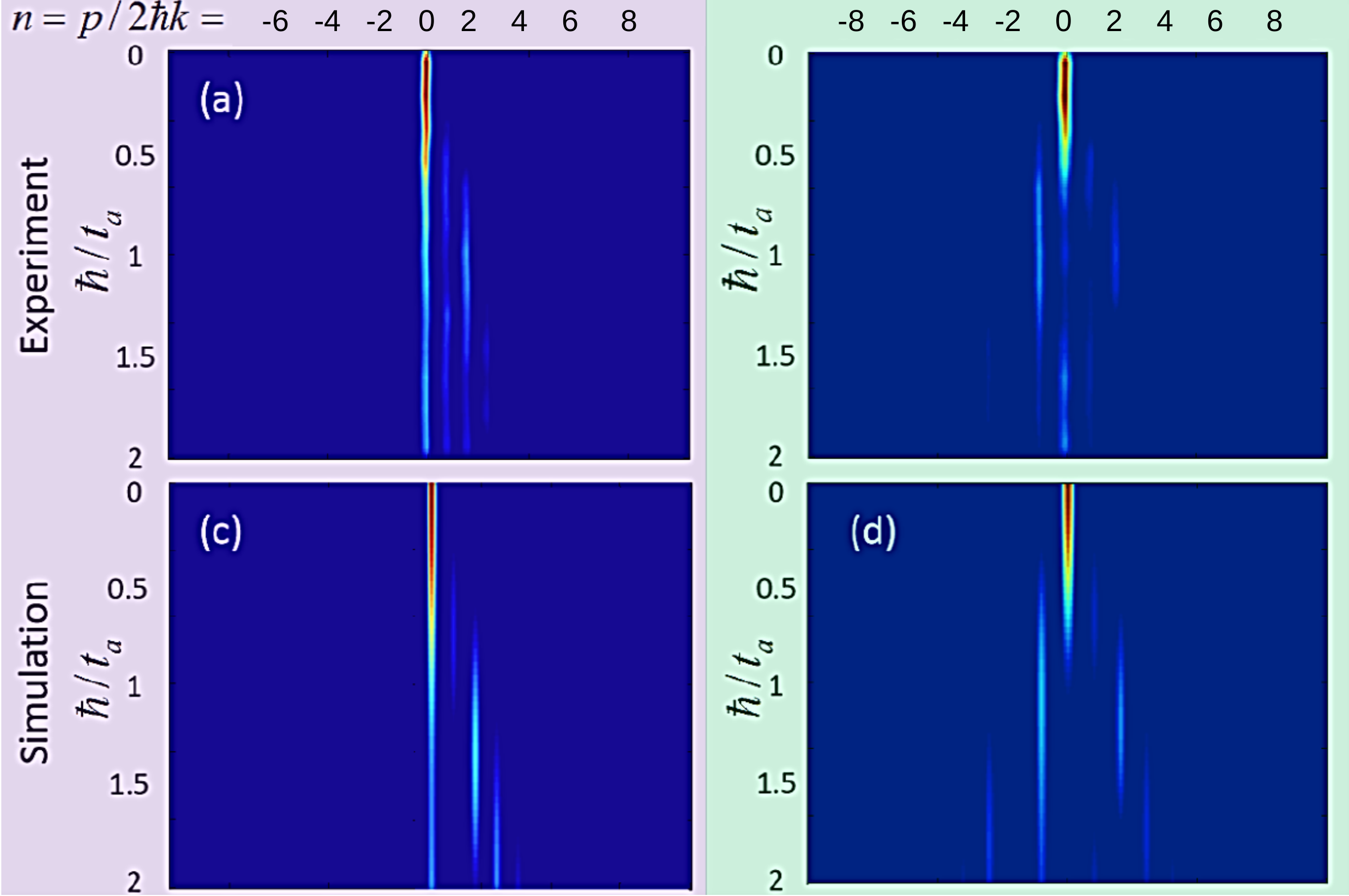}
\caption{\label{figure1}(Color online) Quench dynamics of the SSH4 model. (a) The initial state is prepared at the edge. (b) The theoretical simulations for the edge dynamics.
(c) The initial state is prepared within the bulk. (d) The theoretical simulations for the bulk state. When the initial state is prepared at the edge, the population at the 0 site is always dominant, while for the case of bulk injection, the 0 site population vanishes even for short quench durations. These quench dynamics confirm the boundary correspondence of the bulk topology observed in Fig. 3.}
\end{figure}

\begin{figure}[]
\includegraphics[width= 0.42\textwidth]{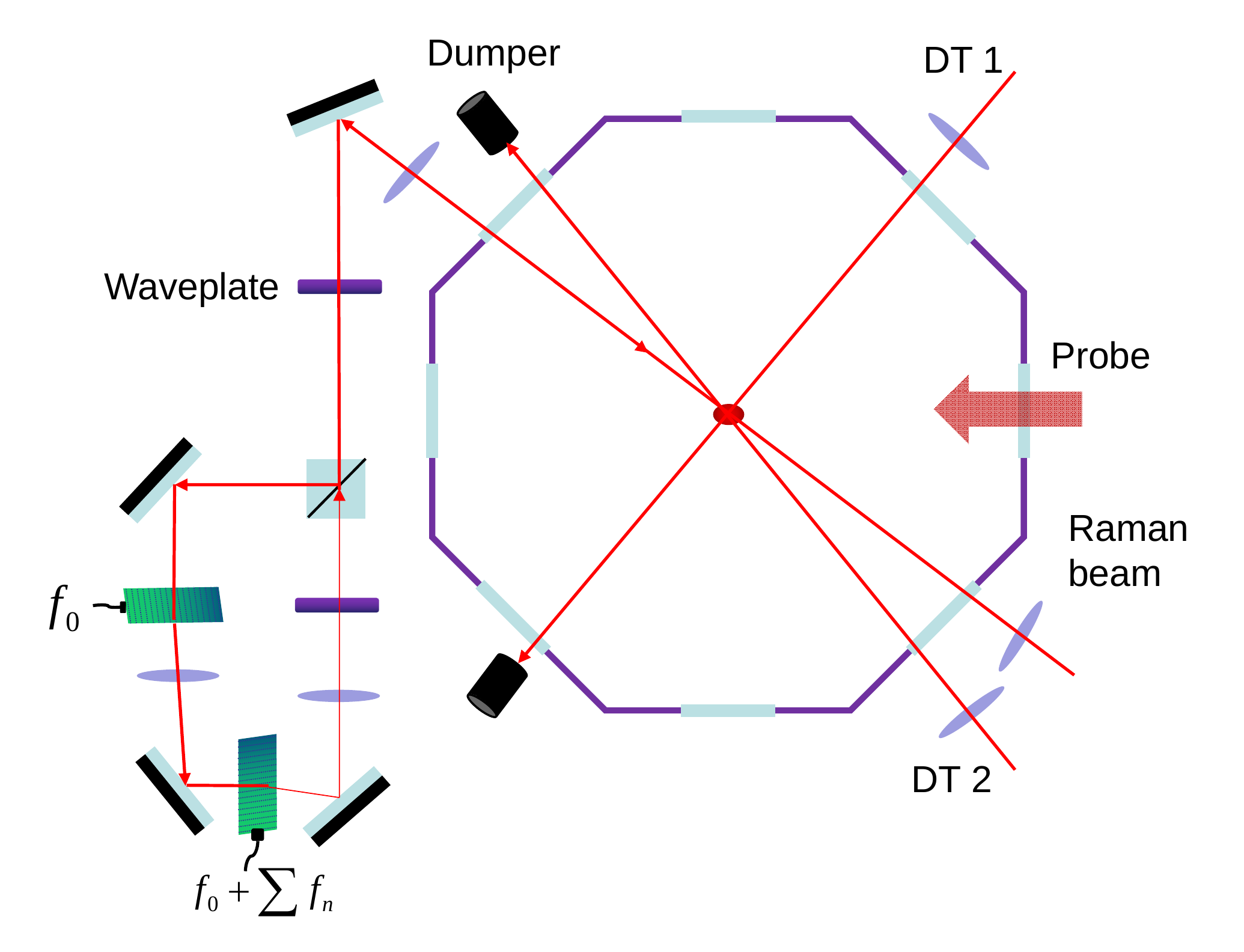}
\caption{\label{figure1}(Color online) (a) The experimental setup. Our BEC is created in a octagonal chamber. Two dipole trap beams (DT 1 and DT 2) form the crossed dipole trap. The Raman incoming beam is about 7 degrees off-axis from the beam DT 2. The incoming Raman beam and the reflected Raman beam are combined with a polarization beam splitter, they are $\sigma +$ and $\sigma -$ at the chamber center. The probe direction is about 45 degree with respect to the Raman direction.}
\end{figure}

\begin{figure}[]
\includegraphics[width= 0.42\textwidth]{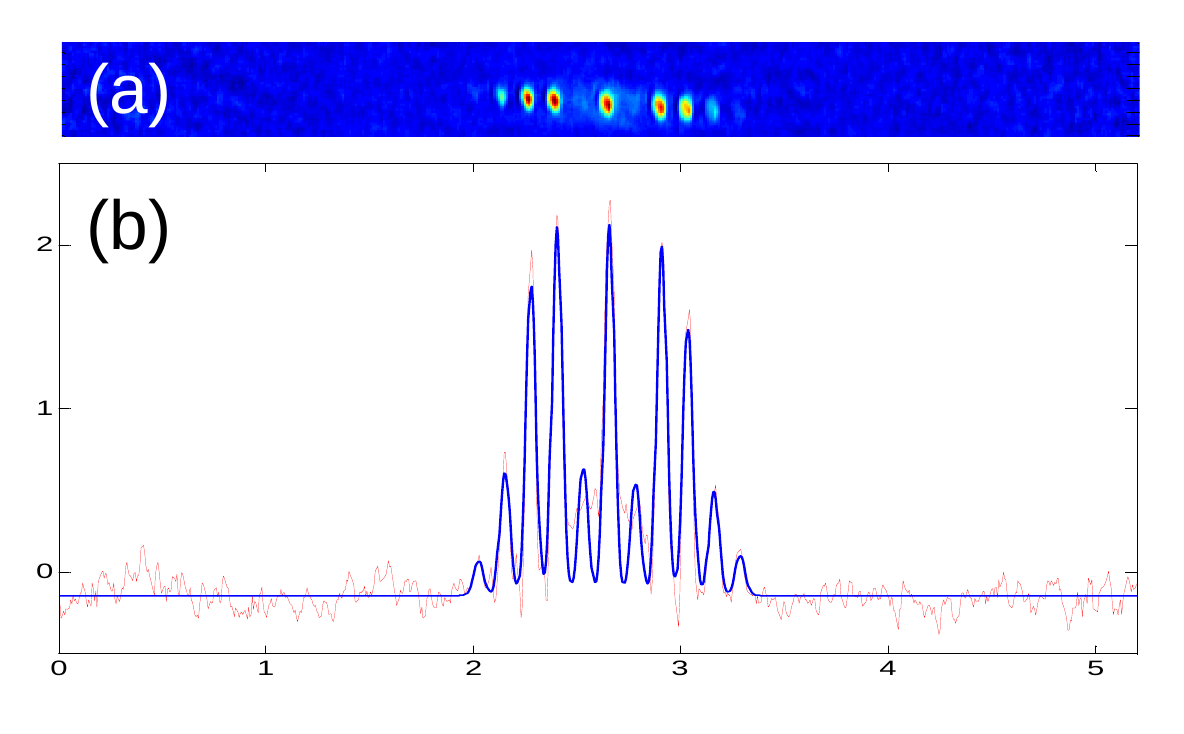}
\caption{\label{figure1}(Color online) Detection method. (a) shows the absorption image of atoms with 20ms free expansion. Atoms in different momentum states occupy different positions after this time-of-flight expansion. These components are are very well separated. (b) shows the multi-Gaussian fitting. Each column of the image shown in (a) is summed, and the size of each Gaussian function is set to be equal. As such, the peak values for each column of the fit orders are proportional to the atom numbers belonging to the different momentum orders.}
\end{figure}

\end{document}